\documentclass[aps,prx,twocolumn,showpacs,superscriptaddress,amsmath,amssymb,amsfonts,floatfix,longbibliography]{revtex4-1}
\usepackage{titlecaps}
\usepackage{graphicx}
\usepackage{placeins}
\usepackage{float}
\usepackage{dcolumn}
\usepackage{bm}
\usepackage{amssymb}
\usepackage[version=4]{mhchem}
\usepackage{microtype}
\newcolumntype{C}[1]{>{\centering\arraybackslash}p{#1}}
\usepackage{xfrac}
\usepackage{gensymb}
\usepackage{xcolor}
\usepackage{multirow}
\usepackage{comment}
\usepackage{times}
\usepackage[varg]{txfonts}
\usepackage{mathrsfs}
\usepackage{amsmath,amssymb,bm,mathtools}
\usepackage{amsfonts}
\usepackage{booktabs}
\usepackage{natbib,hyperref}
\hypersetup{
	colorlinks=true, 
	linkcolor=blue,  
}
\usepackage{soul}
\usepackage{epstopdf}
\usepackage{array}
\usepackage{multirow}
\usepackage{tcolorbox}
\usepackage[percent]{overpic}
\usepackage{bbm}

\usepackage{widetable}
\usepackage{array}
\usepackage{xcolor,colortbl}
\usepackage{lipsum}
\usepackage{tablefootnote}

\newcommand{\ket}[1]{\ensuremath{|#1\rangle}}

\newcommand{\bnzs}{BaNd$_2$ZnS$_5$}
\newcommand{\bczs}{BaCe$_2$ZnS$_5$}

\begin{document}

\title{Quantum entanglement of XY-type spin dimers in Shastry-Sutherland lattice}

\author{Qianli Ma$^*$}
\affiliation{Neutron Scattering Division, Oak Ridge National Laboratory, Oak Ridge, Tennessee, USA, 37831}

\author{Brianna R. Billingsley$^*$}
\affiliation{Department of Physics, University of Arizona, Tucson, AZ, USA, 85721}

\author{Madalynn Marshall}
\affiliation{Department of Chemistry and Biochemistry, Kennesaw State University, Kennesaw, Georgia, USA, 30144}
\affiliation{Neutron Scattering Division, Oak Ridge National Laboratory, Oak Ridge, Tennessee, USA, 37831}

\author{David A. Dahlbom}
\affiliation{Neutron Scattering Division, Oak Ridge National Laboratory, Oak Ridge, Tennessee, USA, 37831}

\author{Yiqing Hao}
\affiliation{Neutron Scattering Division, Oak Ridge National Laboratory, Oak Ridge, Tennessee, USA, 37831}

\author{Daniel M. Pajerowski}
\affiliation{Neutron Scattering Division, Oak Ridge National Laboratory, Oak Ridge, Tennessee, USA, 37831}

\author{Alexander I. Kolesnikov}
\affiliation{Neutron Scattering Division, Oak Ridge National Laboratory, Oak Ridge, Tennessee, USA, 37831}

\author{Xiaojian Bai}
\affiliation{Department of Physics and Astronomy, Louisiana State University , Baton Rouge, Louisiana, USA, 70803}
\affiliation{Neutron Scattering Division, Oak Ridge National Laboratory, Oak Ridge, Tennessee, USA, 37831}

\author{Cristian D. Batista}
\affiliation{Department of Physics and Astronomy, University of Tennessee, Knoxville, Tennessee, USA, 37996}

\author{Tai Kong\textsuperscript{\textdagger}}
\affiliation{Department of Physics, University of Arizona, Tucson, AZ, USA,  85721}
\affiliation{Department of Chemistry and Biochemistry, University of Arizona, Tucson, AZ, USA, 85721}

\author{Huibo Cao\textsuperscript{\textdagger}}
\affiliation{Neutron Scattering Division, Oak Ridge National Laboratory, Oak Ridge, Tennessee, USA, 37831}

\date{\today}

\begin{abstract}

We report a comprehensive study on the origin of the enigmatic disordered ground state within the Shastry-Sutherland lattice, BaCe$_2$ZnS$_5$, at low temperatures. The magnetization and heat capacity data show a lack of magnetic ordering down to 73 mK. We deploy a localized spin dimer model which can accurately reproduce the dynamic structure factor of the neutron data, magnetization and heat capacity data. Remarkably, the intra-dimer exchange interaction shows strong XY-type anisotropy and the ground state of \bczs{} is in an entangled state  $(\ket{\uparrow\uparrow} - \ket{\downarrow\downarrow})/\sqrt{2}$. This is in contrast to the singlet dimer state that is obtained for Heisenberg interactions. These results confirm that \bczs{} is in a quantum paramagnet state consisting of entangled spin dimer states. 

\end{abstract} 

\maketitle
 

Quantum entanglement is an intricate phenomenon that emerges from the principles of quantum mechanics \cite{quantumentanglement2009, entanglementincdm2016}, manifesting itself most strikingly in strongly interacting spin systems~\cite{Gingras2001, kagome2004spin, balents2010spin, gingras2014quantum, highorderSL2020}. Entangled units can be used as  building blocks for studying quantum phase transitions between quantum paramagnets and broken symmetry states or even quantum spin liquid phases~\cite{QSL2000,ruffliquid2007,balents2010spin,gingras2014quantum,bhardwaj2022sleuthing,evanprx}.  
The family of Shastry-Sutherland (SS) materials \cite{Shastry81,mcclarty2017topological,salaba2021,briSSL2023,marshall2023field, AlexSSL2024, PRSSL2024, PRSSLGroundstate2024, magnetocalricSSL2024, Sarah2024} provides a paradigmatic example of the plethora of exotic states of matter that can be induced by applying magnetic field or pressure. 

A key feature of the Shastry-Sutherland (SS) lattice (SSL) is the strong geometric frustration arising from its two-dimensional checkerboard arrangement of orthogonal dimers. This distinctive connectivity results in an exact ground state, characterized by a direct product of singlet states on each dimer, within a finite range of the ratio $J'/J$, where $J$ and $J'$ represent intra-dimer and inter-dimer interactions, respectively. The interplay between the lattice geometry and competing interactions gives rise to a rich phase diagram, which can be compared to the experimentally observed phase diagram of SrCu$_2$(BO$_3$)$_2$, an almost perfect realization of the original SS model \cite{Shastry81, SrSSL1999}.
By applying  pressure to vary the $J'/J$ ratio, SrCu$_2$(BO$_3$)$_2$  evolves from a quantum paramagnetic state consisting of a direct product of singlet on each dimer, for  $J'/J < 0.675$, to an intermediate plaquette phase ($0.675 < J'/J < 0.765$) and eventually settles in a Neel phase ($ J'/J > 0.765$~\cite{zayed20174,SrSSL2020}. 
However, the singlet plaquette phase reported in Ref.~\cite{zayed20174} does not coincide with the plaquette phase that is expected for the SS model~\cite{Miyahara03,Corboz13,Wang18}. 

So far, the known realizations of antiferromagnets on a  SS lattice are limited to Heisenberg interactions realized in SrCu$_2$(BO$_3$)$_2$~\cite{SrSSL1999}, Ising interactions present in rare earth tetraborides RB$_4$ (R = La-Lu)~\cite{Siemensmeyer08,Matas10,Michimura06,Yoshii08} and R$_2$Be$_2$XO$_7$ (R = Rare earch, X = Ge or Si) whose interaction types are still under study\cite{AlexSSL2024, PRSSL2024, PRSSLGroundstate2024, magnetocalricSSL2024, Sarah2024}.
More recently, \bnzs{} has been identified as another realization of Ising-like interactions on the SSL,  exhibiting zero-field magnetic order and field-induced partial disorder~\cite{marshall2023field}.


In this Letter, we report an experimental study of a SSL magnet, \bczs, in which the intra-dimer coupling exhibits a rare XY-type anisotropy. We performed comprehensive thermodynamic characterization and inelastic neutron-scattering (INS) measurement on single-crystal samples of \bczs{} under applied magnetic fields. Our theoretical analysis demonstrates that all experimental data can be accurately described by a localized XY-dimer model. At zero field, this model predicts a ground state characterized by a ferromagnetically entangled wavefunction, $(\ket{\uparrow\uparrow} - \ket{\downarrow\downarrow})/\sqrt{2}$. Furthermore, in the presence of weak inter-dimer couplings, the model predicts a quantum critical point driven by magnetic fields applied within the SSL plane.
These findings introduce a new class of anisotropic interactions into the SSL model, opening avenues for exploring the rich landscape of emergent quantum phases in SSL magnetism.

\begin{figure*}
    \centering
    \includegraphics[width=1\textwidth]{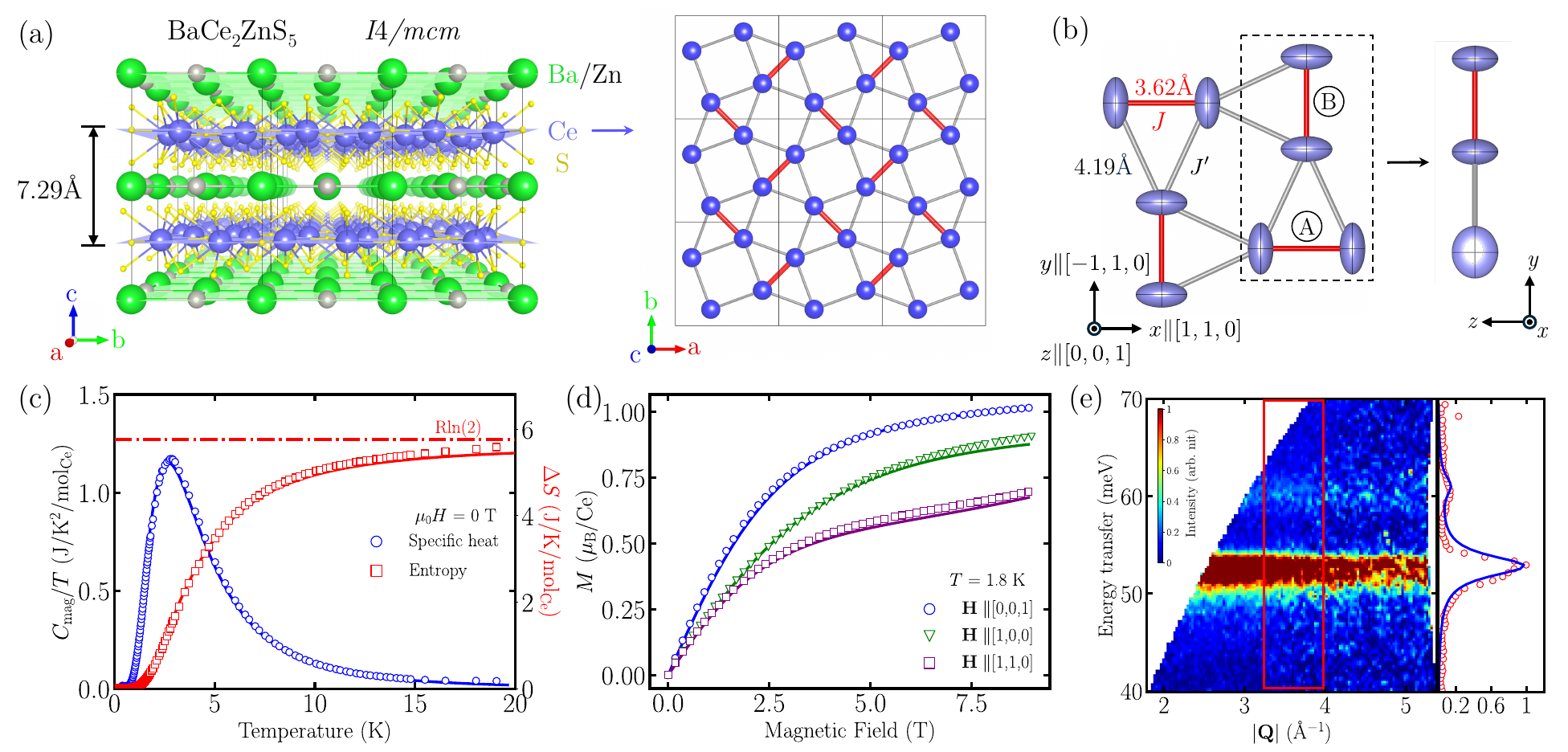}
    \caption{(a) Crystal structure of BaCe$_2$ZnS$_5$ with Barium/Zinc (Ba/Zn), Cerium (Ce), and Sulfur (S) atoms represented by green/gray, blue, and yellow spheres, respectively. The Shastry-Sutherland lattice plane formed by orthogonal dimers of Ce atoms is shown on the right, where nearest-neighbor and next-nearest-neighbor bonds are colored in red and gray, respectively.
(b) Definition of the global coordinate frame used throughout the paper, along with an illustration of the anisotropic $g$-tensor of Ce ions, depicted as ellipsoids. The right panel provides a side view along the $x$-axis.
(c) Specific heat of BaCe$_2$ZnS$_5$ as a function of temperature, measured down to $T\approx70$ mK. The specific heat of the non-magnetic analog BaLa$_2$ZnS$_5$ has been subtracted from the data. The red squares indicate the change in magnetic entropy, which approaches $R\ln(2)$ at high temperatures.
(d) Anisotropic magnetization of BaCe$_2$ZnS$_5$, measured at $T=1.8$\,K along three high-symmetry directions. Solid lines in panels (c) and (d) represent calculations from the localized dimer model.
(e) Powder-averaged inelastic neutron scattering (INS) spectra, showing two excitations at $E=52.57(2)$ meV and $60.84(3)$. The integrated intensities within the red box are displayed on the right, with a fit from the crystal electric field (CEF) analysis shown as the blue curve. }
    \label{fig1}
\end{figure*}

\bczs{} crystallizes in the tetragonal space group $I4/mcm$, hosting SSL layers of magnetic Ce ions separated by layers of non-magnetic cations [Fig.~\ref{fig1}(a)]. The intra- and inter-dimer interactions between Ce ions are mediated via superexchange pathways through sulfur ions. Unlike the sister compound \bnzs{}, which exhibits a long-range magnetic ordering $T_\text{N}=2.9$~K, \bczs{} shows paramagnetic behavior down to $T=1.8$~K in bulk magnetic susceptibility measurements [Fig.~S2 (b)]. The effective magnetic moment extracted from the Curie-Weiss fit of the susceptibility data between 120 and 300~K is 2.52~$\mu_B$, in close agreement with the theoretical value of 2.54~$\mu_B$ for trivalent Ce. Heat capacity data reveal a broad Schottky-like peak centered at $T=3.6$~K, with no additional anomalies observed as low as $T\sim70$~mK [Fig.~\ref{fig1}(c)]. The magnetic entropy change is consistent with the zero-field splitting of the crystal field ground-state doublet, likely due to the dominant intra-dimer interactions. Single-crystal magnetization measured at $T=1.8$~K shows strong dependence on the crystallographic orientations [Fig.\ref{fig1}(d)], suggesting an anisotropic $g$-tensor associated with the crystal field ground-state doublet.

To quantify the magnetic single-ion anisotropy of Ce$^{3+}$, we measured crystal electric field (CEF) excitations using inelastic neutron scattering. Approximately 10 grams of powdered \bczs{} sample was loaded into an aluminum can and mounted on a closed-cycle refrigerator. The measurement was performed on the SEQUOIA time-of-flight spectrometer at the Spallation Neutron Source (SNS), Oak Ridge National Laboratory (ORNL), with an incident energy of $E_\text{i}=80$~meV and a base temperature of $T=5$~K \cite{granroth2010sequoia}. 

The powder-averaged INS spectra, shown in Fig.\ref{fig1}(e), reveal two excitations at $E=52.57(2)$ and $60.84(3)$ meV, resulting from the CEF splitting of the $J=5/2$ manifold of Ce$^{3+}$.
The CEF parameters were extracted from co-fitting of the INS data and bulk magnetic susceptibility using CrysFieldExplorer \cite{MaCEF}. The effective $g$-tensor derived from the ground state wavefunctions exhibits easy-plane anisotropy. The relative orientation of the principal axes with respect to the crystal lattice was determined using the half-polarized neutron diffraction method~\cite{huiboPRL,marshall2023field}, performed on the Dimensional Extreme Magnetic Neutron Diffractometer (DEMAND) at High Flux Isotope Reactor (HFIR), ORNL \cite{cao2018demand}. Further details of the analysis is provided in the Suppl. Sec.~2 and 3. The principal values of anisotropic $g$-tensor for the dimer-A [defined in Fig.~\ref{fig1}(b)] are $g^\text{A}_{xx} = 1.1(2) $, $g^\text{A}_{yy} = 2.6(1)$, and $g^\text{A}_{zz} = 2.2(1)$. The corresponding values for the dimer-B are obtained by a $90^\circ$ rotation along the z-axis, yielding  $g_{xx}^\text{B} =g_{yy}^\text{A} $, $g_{yy}^\text{B} =g_{xx}^\text{A} $ and $g_{zz}^\text{B} =g_{zz}^\text{A} $. 

To understand the microscopic magnetic interactions in \bczs{}, we investigated the low-energy spin excitations using the Cold Neutron Chopper Spectrometer (CNCS) at SNS, ORNL \cite{CNCS}. Single crystals of $\sim\!1.0$\,g were co-aligned in the $hk0$-scattering plane for the INS experiment, with a mosaic spread $\le 5^\circ$. The sample mount was loaded in a superconducting cryomagnet capable of reaching a base temperature of $T=2$\,K and applying a vertical magnetic field up to $\mu_0H=8$\,T. Neutrons with an incident energy of $E_\text{i}=3.32$\,meV were used, providing an elastic energy resolution of FWHM $\sim 0.1$\,meV. INS data were collected at the base temperature by rotating the single-crystal mount along the vertical axis in 1$^\circ$ steps, covering ranges of 240$^\circ$ and 360$^\circ$ under magnetic fields of $\mu_0H=0$ and 4,T, respectively, applied along the $c$-axis.

\begin{figure*}
    \centering
    \includegraphics[width=0.95\textwidth]{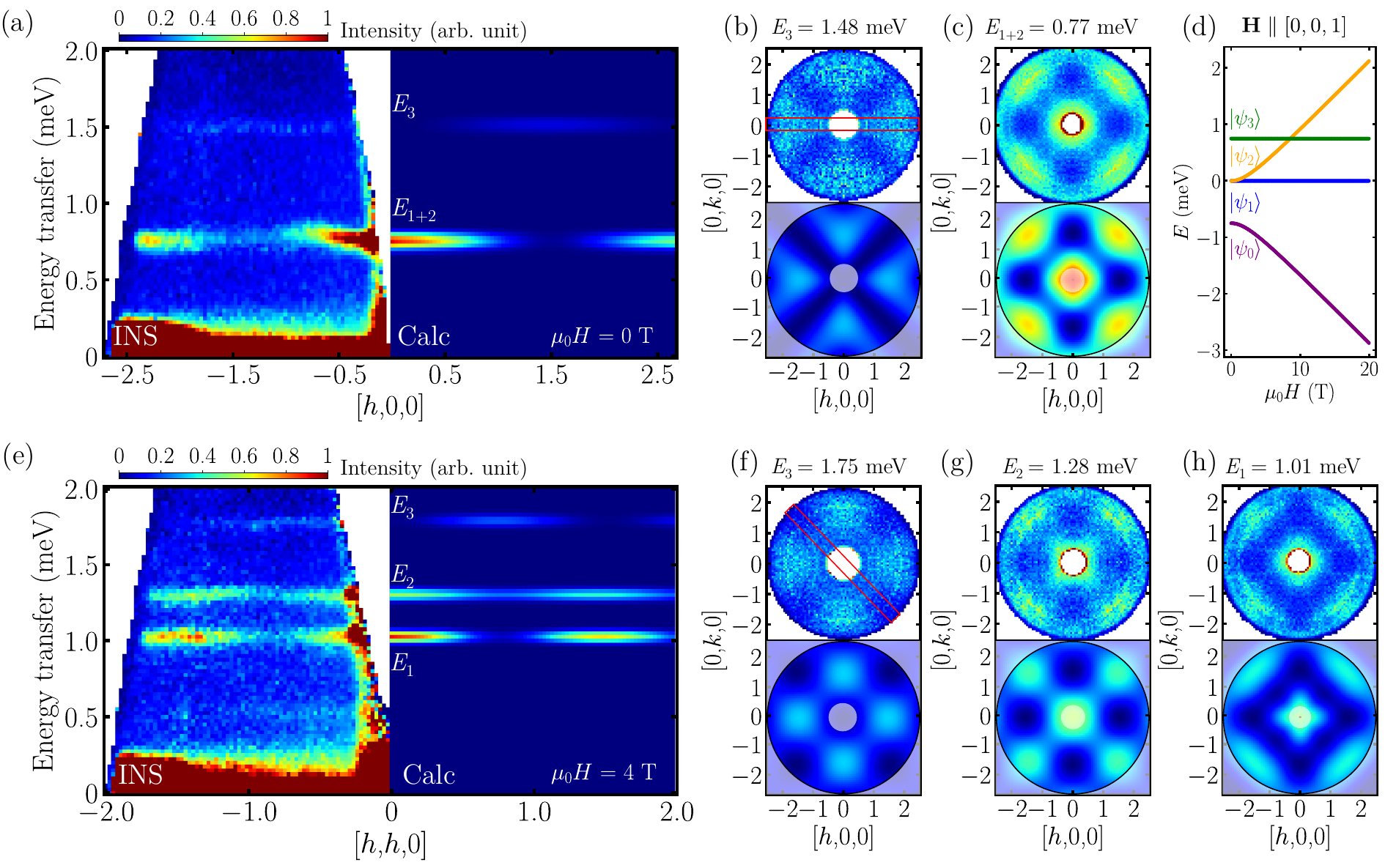}
    \caption{Magnetic excitation spectra of BaCe$_2$ZnS$_5$ at $T=2$\,K measured with an incident neutron energy $E_\text{i}=3.32$\,meV, and comparison with exact diagonalization calculations for the localized dimer model. The data were symmetrized according to the $4/mmm$ point group and integrated over $l=[-1.5,1.5]\,\AA^{-1}$ along the out-of-plane direction. (a, e) Energy-momentum slices of INS data (left) and corresponding simulations (right) performed at $\mu_0H=0$ and 4\,T, respectively. A range of $\pm0.2$\,$\AA^{-1}$ was integrated along the traverse directions. (b-c, f-h) Constant energy slices of INS data at $E_i\pm0.5$\,meV (top) and simulation results (bottom) for selected energies.
(d) Eigenenergy spectrum as a function of field applied along the $[0,0,1]$ direction, where the responses of dimers A and B are identical. 
    }
    \label{neutron}
\end{figure*}

Figure~\ref{neutron} summarizes the low-energy INS data collected at CNCS. At zero field, two quasi-dispersionless bands are observed at $E_{1+2}\approx0.77$\,meV and $E_3\approx1.48$\,meV. Under an applied field of $\mu_0H=4$\,T, the $E_{1+2}$ mode splits into two distinct excitations at $E_1\approx1.01$\,meV and $E_2\approx1.28$\,meV. The total number of observed modes is consistent with the expectation for a dimer of interacting effective spin-$1/2$ degrees of freedom. The quasi-flat nature of these modes indicates that inter-dimer coupling is at a much weaker energy scale. Moreover, the scattering intensity of each mode is strongly modulated in the $hk0$ plane, reflecting the combined effects of dipole transition matrix elements and the crystal sublattice structure.

To perform a quantitative modeling of the observed low-energy excitations, we introduce a minimal Hamiltonian incorporating intra-dimer exchange interactions and Zeeman couplings with external fields:
\begin{eqnarray}
&{\cal H} &= {\cal H}_{\rm intra} + {\cal H}_{\rm Zeeman}\nonumber \\
& &= \sum_{\langle i, j\rangle } {\bf S}^\text{A}_i\cdot{\bf J}^\text{A}\cdot {\bf S}^\text{A}_{j} -  \mu_0\mu_B\sum_{j} {\bf H}\cdot{\bf g}^\text{A}\cdot{\bf S}^\text{A}_j\nonumber\\
& &+\sum_{\langle i, j\rangle } {\bf S}^\text{B}_i\cdot{\bf J}^\text{B}\cdot {\bf S}^\text{B}_{j}  -  \mu_0\mu_B\sum_{j} {\bf H}\cdot{\bf g}^\text{B}\cdot{\bf S}^\text{B}_j
\label{simpHamiltonian}
\end{eqnarray}
where ${\bf S}_i$ is the effective spin-1/2 operator, $\langle i, j \rangle$ denotes the sites $i$ and $j$ within each dimer, and ${\bf J}^\text{A/B}$ represents the intra-dimer exchange tensors for the two sets of orthogonal dimers A and B, as defined in Fig.~\ref{fig1}(b). 
The symmetry-allowed exchange matrix contains three independent parameters:
\begin{equation}\label{J1}
{\bf J}^\text{A/B}=\begin{pmatrix}
J_{xx}^\text{A/B}   & 0   & 0\\
0   & J_{yy}^\text{A/B}   & 0\\
    0    &      0   & J_{zz}^\text{A/B}
\end{pmatrix}\,,
\end{equation}
where $J_{xx}^\text{B} =J_{yy}^\text{A} $, $J_{yy}^\text{B} =J_{xx}^\text{A} $ and $J_{zz}^\text{B} =J_{zz}^\text{A} $.
The dynamical spin structure factor and bulk thermodynamic properties can be calculated directly by diagonalizing the dimer Hamiltonian Eq.~\eqref{simpHamiltonian} in the 4-dimensional Hilbert space spanned by the basis $\{ \ket{\uparrow\uparrow }, \ket{\uparrow\downarrow}, \ket{\downarrow\uparrow},\ket{\downarrow\downarrow}\}$. By varying the three exchange parameters and the three $g$-factors, we obtained an excellent fit to excitation energies of all five observed quasi-flat bands [Fig.~\ref{neutron} (a) and (e)], as well as temperature- and field-dependence of heat capacity and magnetization data [Fig.~\ref{fig1} (c) and (d)]. The best-fit model parameters are:
\begin{eqnarray}
\label{results}
J_{xx}^\text{A} &=\ \ \ 0.06(6),\ \ \ \ g_{xx}^\text{A} &= 1.2(1)\nonumber, \\
J_{yy}^\text{A} &=-1.48(6),\ \ \ \ g_{yy}^\text{A} &= 2.4(2), \\
J_{zz}^\text{A} &= -1.48(6),\ \ \ \ g_{zz}^\text{A} &= 2.1(1).\nonumber
\end{eqnarray} 
The fitted $g$-tensor is consistent with that obtained from the polarized neutron diffraction and the CEF analysis results. See Suppl. Sec. 4. for details of the localized dimer model. 

We further validate the accuracy of our dimer model by calculating the neutron spectral weight distribution for all transitions between the ground state and excited levels, finding excellent agreement with the INS data, as shown in Fig.\ref{neutron}(b)-(c) and (f)-(h). Notably, the model successfully reproduces the key features of the doublet splitting observed between $\mu_0H=0$ and $4$\,T. The circular-shaped intensity profile of the $E_{1+2}$ mode between $[-1, 1]$ along $[h, 0, 0]$ and $[0, k, 0]$ at zero field [Fig.~\ref{neutron}(c)] arises from the superposition of diamond- and square-shaped features associated with the $E_2$ and $E_1$ mode, respectively, which become fully resolved at $\mu_0H=4$\,T, as shown in Fig.~\ref{neutron}(g) and (h).

Two key insights emerge from the dimer model. First, within the estimated uncertainties, the intra-dimer exchange interaction exhibits pure XY-type anisotropy, similar to that of the $g$-tensor. The degeneracy of the $E_{1+2}$ mode is a direct consequence of this XY exchange anisotropy. Second, the ground state at zero field is a ferromagnetically entangled state, $\ket{\psi_0}=(\ket{\uparrow\uparrow} - \ket{\downarrow\downarrow})/\sqrt{2}$, in sharp contrast to that of SrCu$_2$(BO$_3$)$_2$ where the special arrangement of nearest- and next-nearest-neighbor Heisenberg interactions produces a non-magnetic dimer singlet state $ \left(\ket{\uparrow\downarrow}-\ket{\downarrow\uparrow}\right)/\sqrt2$ \cite{SSLlowenergy2001,ssltheory2003}.

Due to the anisotropic nature of spin dimers, the field responses depend strongly on the crystallographic orientation. When the field is applied along the $[0,0,1]$ direction, the responses of dimers A and B are identical. As shown in Fig.~\ref{neutron} (d), the energy eigenstates $\ket{\psi_1}=(\ket{\uparrow\downarrow}+\ket{\downarrow\uparrow})/\sqrt{2}$ and $\ket{\psi_3}=(\ket{\uparrow\downarrow}-\ket{\downarrow\uparrow})/\sqrt{2}$ remain unchanged because they are also eigenstates of $S^z_i+S^z_j$ with eigenvalue zero, while the energy levels $E_0$ and $E_2$ repel each other due to field-induced hybridization between the state $\ket{\psi_2}=(\ket{\uparrow\uparrow}+\ket{\downarrow\downarrow})/\sqrt{2}$ and the ground state $\ket{\psi_0}$. 

\begin{figure}[t] 
    \centering
    \hspace*{-0.2in}
    \includegraphics[width=0.48\textwidth]{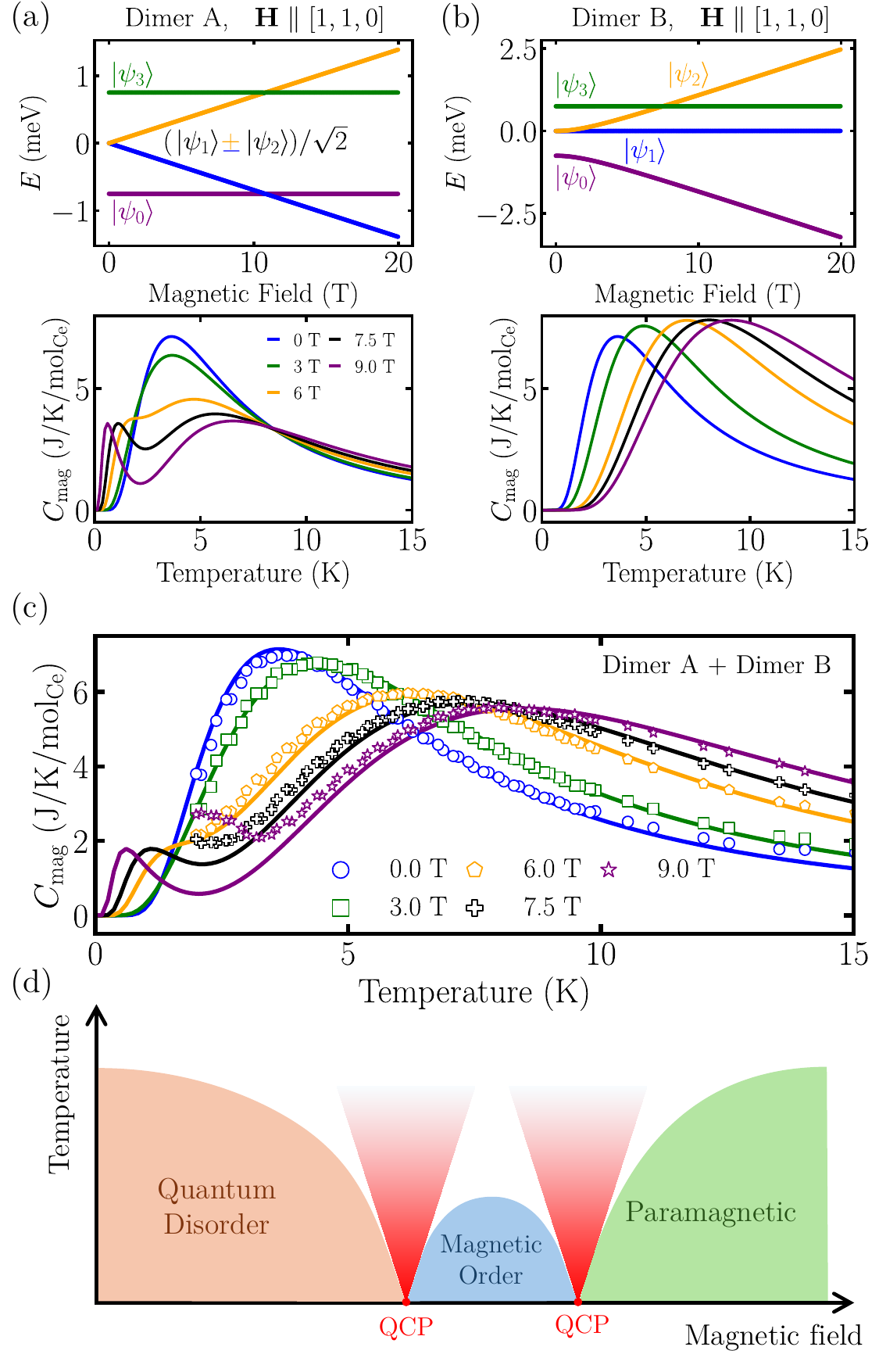}
    \caption{(a,b) Eigenenergy spectrum as a function of field applied along the [1, 1, 0] direction, for dimer A and B, respectively, along with the corresponding simulated specific heat.
(c) Comparison between experimental specific heat data (symbols) and simulations from the localized dimer model (lines). While the model reproduces the temperature and field dependence of the high-temperature peak well, significant discrepancies are evident at low temperatures.
(d) Schematic temperature-field phase diagram illustrating the potential emergence of quantum critical points (QCPs) and an intermediate phase characterized by long-range magnetic order. }
    \label{Cefield}
\end{figure}

When the field is applied along the $[1,1,0]$ direction, the Hamiltonian for dimer A predicts a level crossing between the first excited state $(\ket{\psi_1}-\ket{\psi_2})/\sqrt{2}$ and the ground state $\ket{\psi_0}$ at $\mu_0H \sim 11$\,T [Fig.~\ref{Cefield}(a)]. As a result, the specific heat peak of dimer A splits into two peaks: one shifts to lower temperatures as the gap closes at the level crossing point and then to higher temperatures as the gap reopens, while the other one continuously shifts upward with increasing field.
For dimer B, as the energy levels become further apart with increasing fields [Fig.~\ref{Cefield}(b)], the corresponding heat capacity peak moves monotonically to higher temperatures. By combining the contributions from both dimers, we compare the model predictions with experimental data in Fig.~\ref{Cefield}(c). The calculation shows good agreement for the peak shifting to higher temperatures across all measured fields. However, there is a notable discrepancy for the lower-temperature peak: between $\mu_0H=7.5$ and 9\,T, the experimental data indicate that this peak has already begun to move up in temperatures, implying the reopening of the gap earlier than predicted by the localized dimer model.

This discrepancy suggests a non-trivial effect of inter-dimer couplings near the crossing point. The dimer Hamiltonian can be projected onto the subspace spanned by $\ket{\psi_0}$ and $(\ket{\psi_1}-\ket{\psi_2})/\sqrt{2}$, resulting in an effective low-energy pseudospin-1/2 model. The quasi-degeneracy of these doublets is expected to be lifted by small inter-dimer interactions, giving rise to a magnetically ordered phase, potentially through a quantum phase transition. This transition occurs when the gap becomes comparable to the inter-dimer coupling, which leads to an onset at a smaller field than predicted by the localized dimer model. The ordered phase is expected to form a dome-like structure around the crossing point in the field-temperature phase diagram, as schematically indicated in Fig.~\ref{Cefield}(d). Within this dome, the quasi-particle modes may become significantly more dispersive than those of the quantum disordered phase, driven by the underlying long-range magnetic order.

The form of the local order parameter can be deduced from a general consideration. For dimer A with field applied along the $[1,1,0]$ direction, the $y$- and $z$-component of the total magnetization ${\bf S}^\text{A}_{i}+{\bf S}^\text{A}_{j}$ connects $\ket{\psi_0}$ to $\ket{\psi_1}$ and $\ket{\psi_2}$, respectively. Since the first excited state that becomes degenerate with the ground state at critical point is a superposition $(\ket{\psi_1}-\ket{\psi_2})/\sqrt{2}$, the general form of the local order parameter is expected to be the local magnetization in the $y-z$ plane, as expected from the intra-dimer exchange anisotropy and the single-ion $g$-tensor. Since the uniform field component along the $y$-axis breaks the local U(1) symmetry, we expect an ordered phase only for inter-dimer interactions that favor a {\it non-uniform} ordering of the magnetization of each A-dimer along the $z$ direction.

The ordering wave vector is dictated by the competing inter-dimer interactions and corresponds to the first mode that becomes gapless upon approaching the quantum critical point (QCP) from the low field region. On symmetry grounds, the QCP associated with the onset of the magnetic ordering belongs to the Ising universality class in dimension $D=d+1$, where $d$ is the spatial dimension. Given the Ising-like nature of the quantum phase transition, the gap should reopen on the ordered side of the QCP.

In conclusion, we have conducted comprehensive neutron and bulk magnetic characterizations on both single-crystal and powder samples of \bczs{} material. A minimal spin dimer model was used to resolve the intra-dimer exchange Hamiltonian and the corresponding wave functions for each dimer state, $\psi_i$ (i=0,1,2,3). Notably, this exchange Hamiltonian exhibits strong local XY-anisotropy, and the ground state of \bczs{} forms an entangled state, $\ket{\psi_0} = (\ket{\uparrow\uparrow} - \ket{\downarrow\downarrow})/\sqrt{2}$. Based on the extracted model, the \bczs{} system could enter a field-induced intermediate state through a quantum phase transition around 11 T applied along the [1,1,0] direction. A qualitative phase diagram, shown in Fig.~\ref{Cefield}~(c), describes the temperature-field characteristics of the different states. This places \bczs{}  in closer  resemblance to other frustrated magnetic systems, such as the frustrated pyrochlore Yb$_2$Ti$_2$O$_7$ \cite{Kate2011prx}, the honeycomb $\alpha$-RuCl$_3$ \cite{RuCl32016}, the spin-1 triangular magnet FeI$_2$ \cite{Bai2021} or the ``hidden'' spin-nematic supersolid phase in Na$_2$BaNi(PO$_4$)$_2$ \cite{sheng2023bose}. Strong spin-orbit coupling and crystal electric field effects in many rare-earth and transition metal materials induce pronounced local magnetic anisotropy. Quantum fluctuations arising from the competition between local exchange anisotropy terms have given rise to a diverse array of exotic states of matter~
\cite{Kate2011prx, Bai2021, Ba2CoGe2O72012,NiCl24SC2008, sheng2023bose}. This makes \bczs{} a promising candidate to test the combined effects of competing frustrated inter-dimer interactions and local exchange anisotropy as it is the first Shastry-Sutherland material in the BaR$_2$ZnX$_5$ family (R = rare earth, X = S or O) where the anisotropic exchange Hamiltonian has been fully resolved.

\vspace{0.3cm}

The research at Oak Ridge National Laboratory (ORNL) was supported by the U.S. Department of Energy (DOE), Office of Science, Office of Basic Energy Sciences, Early Career Research Program Award KC0402020, under Contract DE-AC05-00OR22725. The work at the University of Arizona was supported by the U.S. Department of Energy (DOE), Office of Science, Basic Energy Sciences (BES) under Award DE-SC0025301. B.R.B acknowledge support from the National Science Foundation under Award DGE-2137419. The work at Louisiana State University was supported by the U.S. Department of Energy (DOE), Office of Science, Basic Energy Sciences (BES) under Award DE-SC0025426. This research used resources at the Spallation Neutron Source and High Flux Isotope Reactor, a DOE Office of Science User Facility operated by ORNL. This research was partially supported by the National Science Foundation Materials Research Science and Engineering Center program through the UT Knoxville Center for Advanced Materials and Manufacturing (DMR-2309083).

\end{document}


\title{Supplemental Information:\\ Quantum entanglement of XY-type spin dimers in Shastry-Sutherland lattice}

\author{Qianli Ma$^*$}
\affiliation{Neutron Scattering Division, Oak Ridge National Laboratory, Oak Ridge, Tennessee, USA, 37831}

\author{Brianna R. Billingsley$^*$}
\affiliation{Department of Physics, University of Arizona, Tucson, AZ, USA, 85721}

\author{Madalynn Marshall}
\affiliation{Department of Chemistry and Biochemistry, Kennesaw State University, Kennesaw, Georgia, USA, 30144}
\affiliation{Neutron Scattering Division, Oak Ridge National Laboratory, Oak Ridge, Tennessee, USA, 37831}

\author{David A. Dahlbom}
\affiliation{Neutron Scattering Division, Oak Ridge National Laboratory, Oak Ridge, Tennessee, USA, 37831}

\author{Yiqing Hao}
\affiliation{Neutron Scattering Division, Oak Ridge National Laboratory, Oak Ridge, Tennessee, USA, 37831}

\author{Daniel M. Pajerowski}
\affiliation{Neutron Scattering Division, Oak Ridge National Laboratory, Oak Ridge, Tennessee, USA, 37831}

\author{Alexander I. Kolesnikov}
\affiliation{Neutron Scattering Division, Oak Ridge National Laboratory, Oak Ridge, Tennessee, USA, 37831}

\author{Xiaojian Bai}
\affiliation{Department of Physics and Astronomy, Louisiana State University , Baton Rouge, Louisiana, USA, 70803}
\affiliation{Neutron Scattering Division, Oak Ridge National Laboratory, Oak Ridge, Tennessee, USA, 37831}

\author{Cristian D. Batista}
\affiliation{Department of Physics and Astronomy, University of Tennessee, Knoxville, Tennessee, USA, 37996}

\author{Tai Kong\textsuperscript{\textdagger}}
\affiliation{Department of Physics, University of Arizona, Tucson, AZ, USA,  85721}
\affiliation{Department of Chemistry and Biochemistry, University of Arizona, Tucson, AZ, USA, 85721}

\author{Huibo Cao\textsuperscript{\textdagger}}
\affiliation{Neutron Scattering Division, Oak Ridge National Laboratory, Oak Ridge, Tennessee, USA, 37831}

\date{\today}

\maketitle

\section{Material Synthesis and Characterization}

Single crystals of BaCe$_2$ZnS$_5$ were synthesized using the high-temperature solution growth method identical to the procedure of \bnzs \cite{briSSL2023}. Starting elements of barium pieces (Alfa Aesar, 99.2\%), cerium pieces (Alfa Aesar, \%), zinc shot (Alfa Aesar, 99.999\%), and sulfur pieces (Alfa Aesar, 99.999\%) were packed in an alumina Canfield Crucible Set and sealed in a silica tube under vacuum with molar ratios of Ba:Ce:Zn:S = 5:4:2:19. Air-sensitive barium and cerium were handled and packed in an argon glovebox. Once sealed, the ampoule was first heated to 430~$^\circ$C over 3 hours, dwelled for 10 hours, then heated to 850~$^\circ$C over 12 hours and dwelled there for 5 hours. The ampoule was further heated to 1060~$^\circ$C over 6 hours and dwelled at 10 hours before slowly cooling to 750~$^\circ$C for decanting. Slow heating during this process is essential to prevent explosion due to the large amounts of sulfur vapor pressure present and to protect the crucible from being attacked by the barium and cerium. Risk of explosion is still present during the decanting stage as significant sulfur vapor pressure inside the silica tube can be visually observed. Cubic-like single crystals of BaCe$_2$ZnS$_5$ can be collected from the crucible after decanting. Ba-S surface impurity is removed with distilled water. Cleaned crystals measure about 1-2 mm in length and can vary in mass from 1~mg to 15~mg.

\begin{figure}
    \centering
    \hspace*{-0.0in}
    \includegraphics[width=0.5\textwidth]{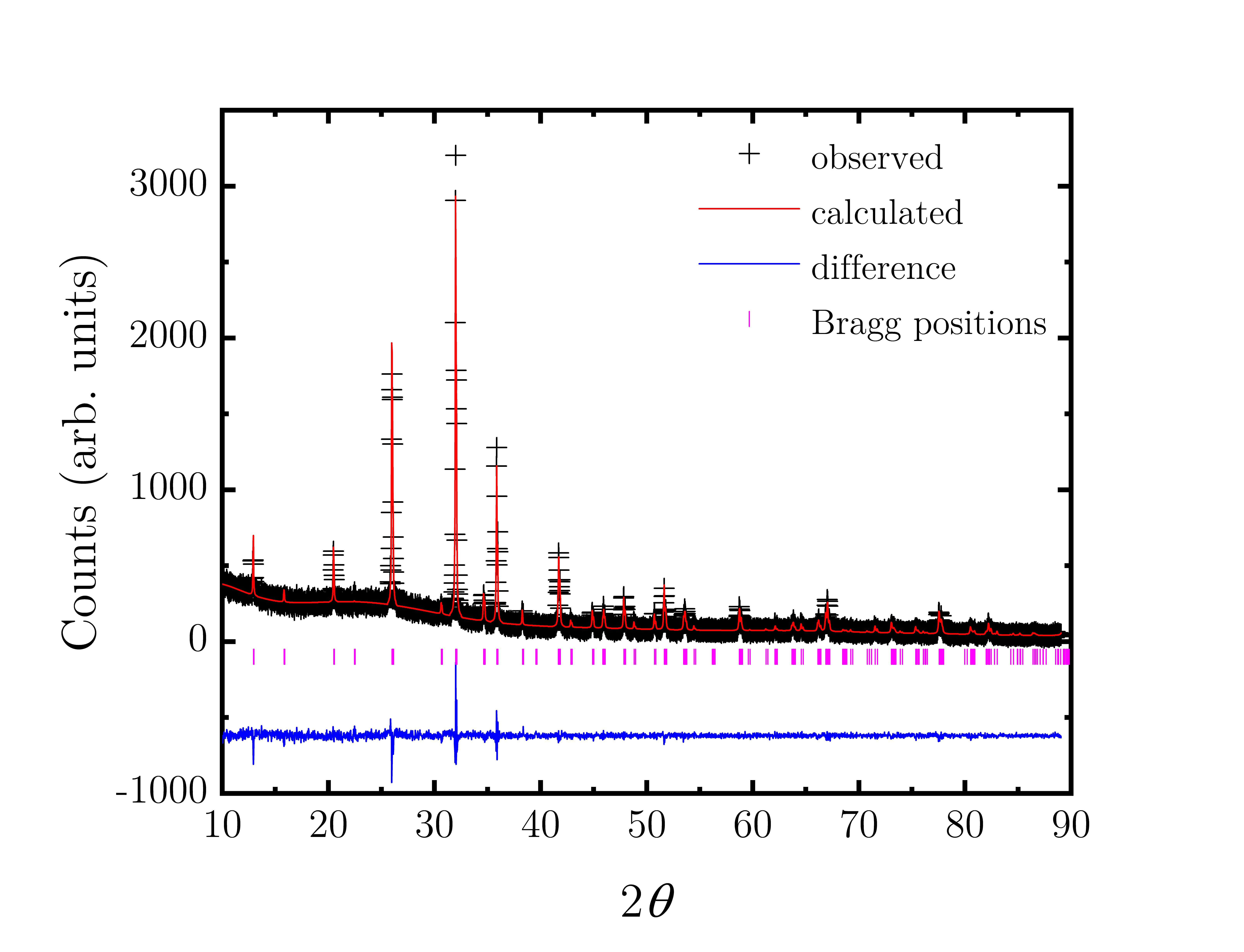}
    \caption{Calculated (line) and measured (crosses) powder x-ray diffraction data plotted as intensity counts vs 2$\theta$. The fitting shows minimal residual (blue line) agreeing well with the reported structure in Fig.~1 in the main manuscript.}
    \label{PXRD}
\end{figure}

After collecting and grinding a selection of BaCe$_2$ZnS$_5$ single crystals into a fine powder and spread onto a vacuum greased glass slide, room-temperature powder x-ray diffraction (XRD) was measured using a Bruker D8 Discover diffractometer with a Cu K$_\alpha$ radiation ($\lambda$ = 1.5406 Å) shown in Fig.~\ref{PXRD}. The powder XRD data was analyzed using Rietveld refinement. Crystalline orientation was determined by single crystal x-ray diffraction on the same diffractometer and measurement of the subsequent single crystalline diffraction peaks. Powder XRD results indicate that the grown crystals are pure BaCe$_2$ZnS$_5$, and single crystal XRD shows that the crystal facets are [0,0,1] and [1,1,0].

With the crystal orientation identified, anisotropic magnetization and specific heat data were measured using a Quantum Design Physical Property Measurement System (Dynacool; 1.8 to 300 K, 0 to 90 kOe). Magnetization was measured using the vibrating sample magnetometer function (VSM). Single crystalline samples were manually mounted on a silica sample holder in preferred orientation with GE varnish. A QD dilution refrigerator insert was used to measure specific heat data down to $\sim$70~mK. The specific heat of a BaLa$_2$ZnS$_5$ single crystal was measured to serve as the nonmagnetic lattice contribution for the estimation of magnetic entropy in BaCe$_2$ZnS$_5$ as done previously with BaNd$_2$ZnS$_5$. \cite{briSSL2023}

\section{Crystal Electric Field Analysis}
As discussed in the main manuscript, single-ion anisotropy was investigated by performing crystal electric field (CEF) analysis on $\sim$ 10 grams of powder \bczs sample on TOF spectrometer SEQUOIA at SNS, ORNL \cite{granroth2010sequoia}. Ce$^{3+}$ occupies atomic site with $C_{2\nu}$ point symmetry which allows nine non-zero CEF parameters. This is further reduced to 5 parameters due to the 6th order of Stevens factor $\gamma$ = 0. We construct the CEF Hamiltonian using the allowed Stevens operators as:
\begin{eqnarray}
\label{results}
    {\cal H_{\rm CEF}} = B_2^0{O}_2^0 +B_2^2{O}_2^2 + B_4^0{O}_4^0 + B_4^2{O}_4^2 + B_4^4{O}_4^4.
\end{eqnarray}

\begin{figure}[t] 
    \centering
    \hspace*{-0.2in}
    \includegraphics[width=0.45\textwidth]{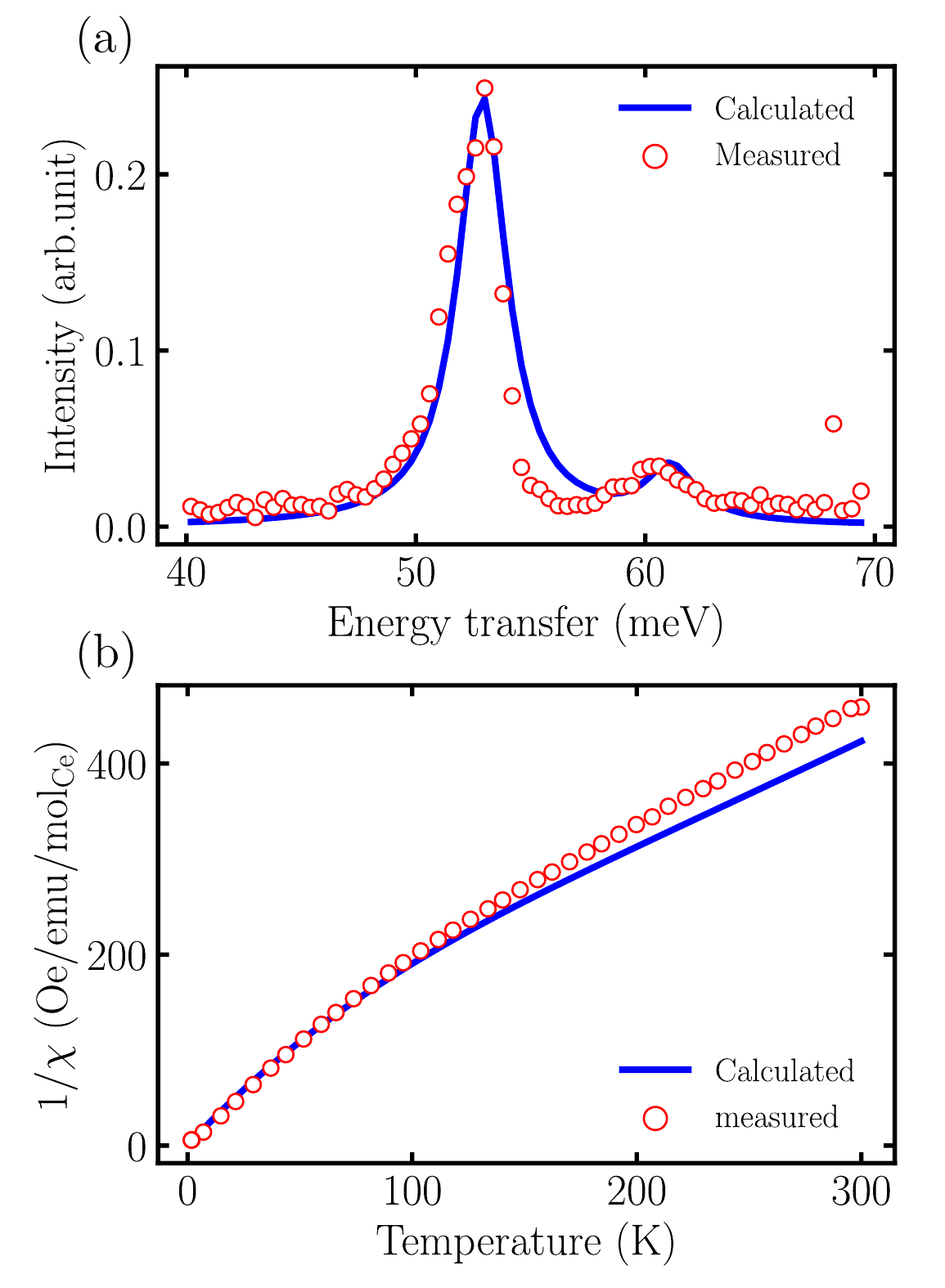}
    \caption{(a) The calculated and measured integrated intensities of constant $|\bf Q|$ cut from 3.4 to 4 $\AA^{-1}$. (b) The calculated and measured inverse magnetic susceptibility of \bczs{.}}
    \label{SI-CEF}
\end{figure}

The powder averaged neutron spectra has been shown in Fig.1 (e) in the main manuscript. Ce$^{3+}$ has a total angular momentum number $J$ = 5/2, resulting in 2$J$+1 = 6-fold degeneracy. The degeneracy is lifted due to the CEF effect into 3 doublets. The integrated intensities along energy transfer between $|\bf Q|$ from 3.4 - 4 $\AA^{-1}$ was used to identify the CEF transitions and shown in Fig.~\ref{SI-CEF} (a). Two peaks at 52.57 (2) meV and 60.84(3) meV are identified as the crystal field excitations from the ground state doublet to the first and second excited states. CrysFieldExplorer was used to perform a global fitting on the CEF Hamiltonian to capture the experimentally measured neutron energy levels, relative intensity and inverse magnetic susceptibility data \cite{MaCEF}. The refined CEF parameters are $B_2^0$ = $-1.983(3)$ meV, $B_2^2$ = $-3.291(3)$ meV, $B_4^0$ = $0.0051(5)$ meV, $B_4^2$ = $-0.122(5)$ meV and $B_4^4$ = $-0.787(5)$ meV. The displayed parameters have been multiplied by the Stevens factors for Ce$^{3+}$ \cite{jensen1991rare}. The uncertainties are estimated by varying the parameters to produce an observable effect. The results are shown as solid lines in Fig.~\ref{SI-CEF}.

The fitted neutron energy and relative intensities are tabulated in Table.~S\ref{CEF-data} while the corresponding eigenvectors in Table.~S\ref{CEF-vec}.

\begin{table}[]
    \begin{tabular}{|c|c|c|}
    \hline
                      &  Obs   & Calc\\
    \hline
        $E_1$ (meV) & 52.57 (2) & 52.89 (2)\\
    \hline
        $E_2$ (meV) & 60.84 (3) & 61.10 (5)\\
    \hline
        $I_2$/$I_1$ (arb.units) & 0.125 (3) & 0.125\\
    \hline
    \end{tabular}
    \caption{Comparison between the observed and calculated energies and relative intensity of the CEF excitations in \bczs{.}}
    \label{CEF-data}
\end{table} 

\begin{table}[]
    \begin{tabular}{|c|c|c|c|c|c|c|}
    \hline
    \backslashbox[1mm]{$\ket{m_J}$}{$E$(meV)}
        & 0 & 0 & 52.89 & 52.89 & 61.10 & 61.10  \\
    \hline
         \ket{-5/2}  & 0.81 & 0    & $-0.58$ & $-$0.01 & $-$0.08 &   0\\
    \hline
         \ket{-3/2}  & 0    & 0.50 & $-$0.02 &  0.75 &  0.02 &	0.42\\
    \hline
         \ket{-1/2}  & 0.31 & 0    &  0.30 &  0.01 &  0.90 &	$-$0.04\\
    \hline
         \ket{1/2}   & 0    & 0.31 & $-$0.01 &  0.30 & $-$0.04 &	$-$0.90\\
    \hline
         \ket{3/2}   & 0.50 & 0    &  0.75 &  0.02 & $-$0.42 &	0.02\\
    \hline
         \ket{5/2}   & 0    & 0.81 &  0.01 & $-$0.58 &   0   &	0.08\\
    \hline
    \end{tabular}
    \caption{Eigenvectors associated with the CEF Hamiltonian of \bczs{.} The first row displays the calculated crystal field spectrum. The corresponding eigenvectors are given in each column in $m_J$ basis.}
    \label{CEF-vec}
\end{table} 

\section{Polarized Neutron Diffraction Measurements}

Polarized single crystal neutron diffraction measurements were performed on the HB-3A Dimensional Extreme Magnetic Neutron Diffractometer (DEMAND) at the High Flux Isotope Reactor at Oak Ridge National Laboratory \cite{cao2018demand}. A polarized neutron beam of 1.542 Å from a bent Si-220 monochromator was used with a calibrated neutron polarization of 78$\%$ \cite{chakoumakos2011four}. Two experiments were performed one with a fixed field of 0.6 T along [1, $-$1, 0] and another with a fixed field of 0.48 T along [0,0,1] using a permanent magnet set and loaded in a closed-cycle refrigerator. A total of 16 flipping ratios were measured at 5 K and were analyzed using the CrysPy software \cite{CrysPy}. 

\begin{figure}
    \centering
    \hspace*{-0.2in}
    \includegraphics[width=0.55\textwidth]{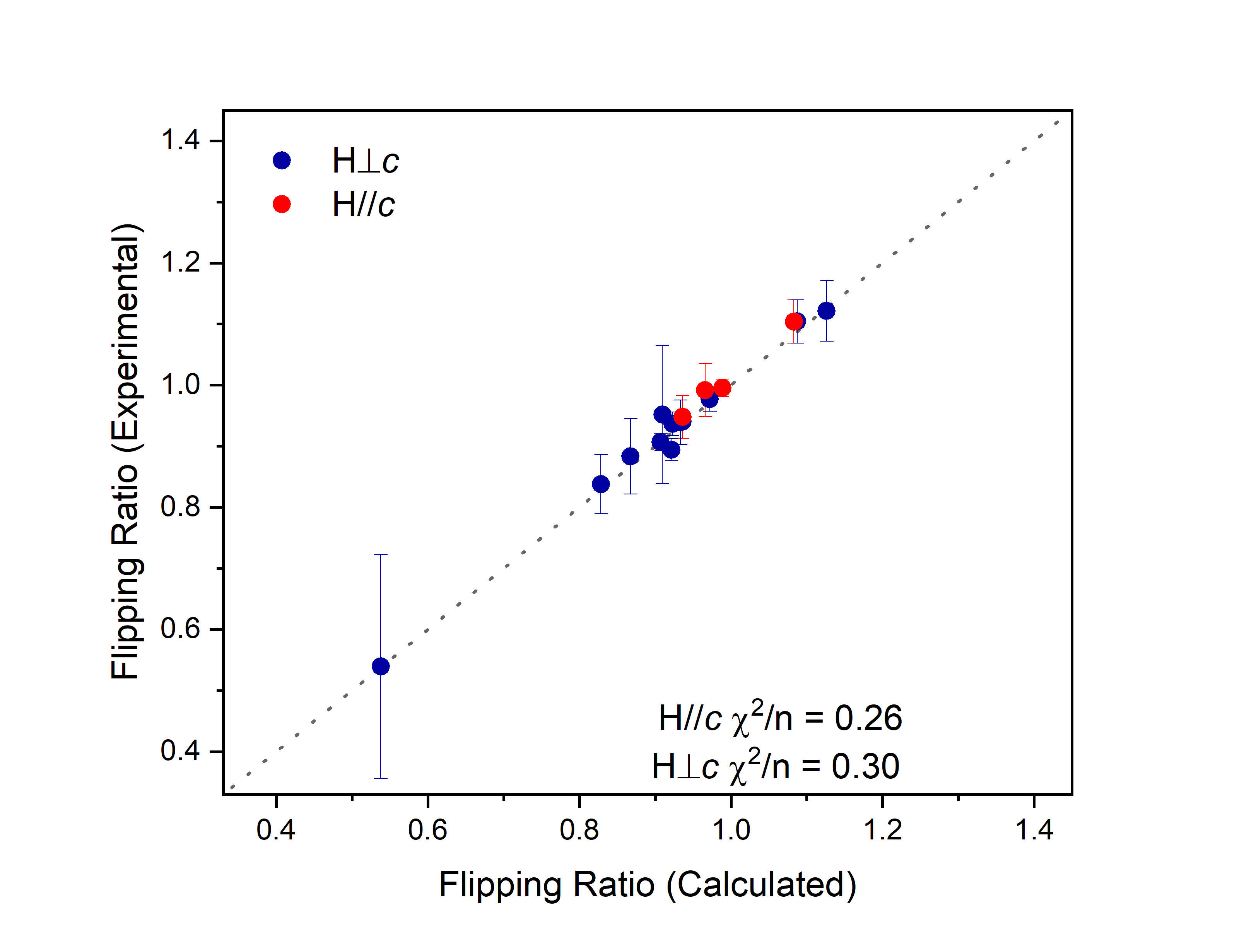}
    \caption{The fitted and calculated flipping ratios for H$\parallel$  (red) and H$\perp$c (blue) at 0.6 T at 5 K.}
    \label{polarized}
\end{figure}

The paramagnetic state of \bczs was investigated using polarized neutron diffraction by measuring the local magnetic susceptibility tensor of the Ce$^{3+}$ spins. Contrary to BaNd$_2$ZnS$_5$ \cite{marshall2023field}, a strong anisotropy was not observed in the bulk measurements of BaCe$_2$ZnS$_5$. To first determine the in-plane magnetic anisotropy, we measured reflections in spin-up and spin-down neutron channels with a magnetic field along [1, $-$1, 0]. Similar to the reported BaNd$_2$ZnS$_5$\cite{marshall2023field}, the principal axes of the ellipsoid are along the [1, 1, 0], [1, $-$1, 0] and [0, 0, 1] directions given the local symmetry from the 8$h$ site with space group $I4/mcm$ of the Ce atomic site in BaCe$_2$ZnS$_5$. The two free susceptibility tensor parameters in-plane were refined with the obtained 12 good-quality flipping ratios. An in-plane Ising-spin nature with moments
orthogonal to the dimer bond was observed with lengths $\chi_{\parallel}= 0.23(2)\mu_B/T$ and $\chi_{\perp}= 0.03(2) \mu_B/T$. To determine the out-of-plane magnetic anisotropy, we measured the local magnetic susceptibility tensor of the Ce$^{3+}$ spins with a magnetic field along the c axis. Here only the out-of-plane susceptibility tensor was refined with the 4 good-quality flipping ratios to obtain the length $\chi = 0.137(31) \mu_B/T$. These results suggests strong easy-plane anisotropy in the directions perpendicular to the intra-dimer bonds, shown in Fig.~1 (b) in the main manuscript.

\section{Localized Dimer Model}
In the main manuscript, a localized dimer model was constructed to describe the single crystal inelastic neutron data in conjunction with magnetization and specific heat measurements. The Hamiltonian of the localized dimer model is shown in Eq.~1 of the main manuscript. It considers intra-dimer interactions of dimer A, dimer B and their respective Zeeman terms. No inter-dimer interaction is considered in this model. 

Symmetry analysis limits the allowed matrix components of the intra-dimer exchange interactions to three independent terms. In the global frame defined in the main manuscript where the $x\parallel$[1,1,0], $y\parallel$[-1,1,0] and $c\parallel$[0,0,1], the exchange matrix for the intra-dimer bond takes on the diagonalized form of Eq.~2 in main manuscript. Note that dimer A and dimer B are connected through a 90$\degree$ rotation along c-axis, ${\bf J}^A = R^{-1}_z(90) \cdot {\bf J}^B\cdot R_z(90)$. This transformation indicates the relation $J^B_{xx} = J^A_{yy}$, $J^B_{yy}=J^A_{xx}$ and $J^B_{zz}=J^A_{zz}$ as noted in the main manuscript.

The matrix representation of the spin operators can be constructed by taking Kronecker product between Pauli matrices and the identity matrix as:
\begin{eqnarray}
S^{\alpha}_{m} &=& \dfrac{1}{2} (\sigma_{\alpha} \otimes \mathbf{1})\nonumber \\
S^{\beta}_{n} &=& \dfrac{1}{2} (\mathbf{1} \otimes \sigma_{\beta})
\label{spins}
\end{eqnarray}
where $\alpha,\beta$ = $x,y,z$. 



Using the matrix representation of the spin operators and exchange matrices in Eq.~2, we can explicitly diagonalize Eq.~1 in the main manuscript and analytically resolve the eigenvalues and eigenfunctions of the dimer model. The resultant eigenvalues and eigenfucntions are:

\begin{eqnarray}
\label{Cewavefun0T_0}
E_0 &=&\dfrac{1}{4} (-J^{A/B}_{xx} + J^{A/B}_{yy} + J^{A/B}_{zz}) \nonumber \\
\label{Cewavefun0T_1}
E_1 &=&\dfrac{1}{4} (J^{A/B}_{xx} + J^{A/B}_{yy} - J^{A/B}_{zz}) \nonumber \\
\label{Cewavefun0T_2}
E_2 &=&\dfrac{1}{4} (J^{A/B}_{xx} - J^{A/B}_{yy} + J^{A/B}_{zz}) \nonumber \\
\label{SI-E}
E_3 &=&\dfrac{1}{4} (-J^{A/B}_{xx} - J^{A/B}_{yy} - J^{A/B}_{zz}).
\end{eqnarray}

\begin{eqnarray}
\label{Cewavefun0T_0}
E_0:\ket{\psi_0} &=&\dfrac{1}{\sqrt{2}} (\ket{\uparrow\uparrow}-\ket{\downarrow\downarrow}) \nonumber \\
\label{Cewavefun0T_1}
E_1:\ket{\psi_1} &=&\dfrac{1}{\sqrt{2}} (\ket{\uparrow\downarrow}+\ket{\downarrow\uparrow}) \nonumber \\
\label{Cewavefun0T_2}
E_2:\ket{\psi_2} &=&\dfrac{1}{\sqrt{2}} (\ket{\uparrow\uparrow}+\ket{\downarrow\downarrow}) \nonumber \\
\label{SI-eig}
E_3:\ket{\psi_3} &=&\dfrac{1}{\sqrt{2}} (\ket{\uparrow\downarrow}-\ket{\downarrow\uparrow}).
\end{eqnarray}

The ground state and the excited states can be determined by fitting the eigenvalues with neutron data shown in Fig.~2 from the main manuscript or Sec. 5 of the SI. Zero-field and 4 T neutron data captures 5 transitions in total, combined with magnetization and specific heat data we can accurately parameterize the three independent matrix components $J^{A/B}_{xx}$, $J^{A/B}_{yy}$ and $J^{A/B}_{zz}$. The general formula that computes the cross section of the neutron dynamic structure factors between two transitions is given in Ref.\cite{crosssection}:

\begin{eqnarray}
 \dfrac{\partial^2\sigma}{\partial\Omega\ \partial E}&=& A\sum_{i, j} \exp{-\dfrac{E(\psi)}{kT}} \sum_{\alpha,\beta}(\delta_{\alpha,\beta} - \dfrac{Q_\alpha Q_\beta}{Q^2}) \times \nonumber \\
 &&\sum_{m,n}F^*_m({\bf Q})F^*_n({\bf Q})\exp\left(\text{i}{\bf Q}({{\bf R_m} - {\bf R_n}})\right)\bra{\psi_i}M^\alpha_m\ket{\psi_j} \times \nonumber \\
 &&\bra{\psi_j}M^\beta_n\ket{\psi_i}\delta(E + E_i - E_j)
\label{cross-section}
\end{eqnarray}
where $\alpha, \beta= x,y,z$. m and n number the magnetic ions within a dimer unit (in our case, both dimer spins are Ce$^{3+}$). $F_{m,n}(\Vec{Q})$ is the magnetic form factor for Ce. ${\bf R}_m$ are the position vectors of the Ce$^{3+}$ in the dimer unit. The quantity $A=\gamma e^2 k' / m_e c^2 k \exp(-2W)$, which is made up from a constant and Debye-Waller factor. 

SI-Eq.~\ref{cross-section}, allows explicit evaluations of the matrix components $\bra{\psi}\hat{S^\alpha_m}\ket{\psi'}$, $\bra{\psi'}\hat{S^\beta_n}\ket{\psi}$ and the interference factor $\exp[\text{i}{\bf Q}({{\bf R_m} - {\bf R_n}})]$. These provide strong modulation on the in-plane dynamic structure factor. Matching the simulated results with experimental data provides strong constraints of the wavefunctions for each levels. Our calculation shows excellent agreement with neutron data as shown in Fig.~2 in main manuscript, therefore, allowing us to determine the wavefunctions for each level as shown in SI-Eq.~\ref{SI-eig}. As can be seen, the ground state is in a ferromagnetically entangled state $\ket{\psi_0} = (\ket{\uparrow\uparrow}-\ket{\downarrow\downarrow})/\sqrt{2}$.

\subsection{Calculation of Magnetization}

Magnetization along high symmetry directions shown in Fig.~1 (d) provides further constraints on exchange parameters and anisotropy $g$-tensor components. The formulate to calculate magnetization is given as:

\begin{eqnarray}
 {\bf M}({\bf H},T) = \frac{1}{N_\text{spin}}\sum_m\dfrac{1}{Z}\sum_{i} \bra{\psi_i({\bf H})}{\bf g}_m\cdot{\bf S}_{m}\ket{\psi_i({\bf H})}e^{-E_i/k_\text{B}T}
\label{mag}
\end{eqnarray}
where $Z=\sum_i \exp{(-E_i/k_\text{B}T)}$ is the partition function, $i$ labels eigenstates of the localized dimer model and $m$ labels spin in the unit cell. The calculated results are shown as solid lines in Fig.~1 (d) in the main text.

\subsection{Calculation of Specific Heat}

Specific heat calculation only involves excited energy levels calculated from the localized dimer model. It provides further constraints on the overall fitting. The formulate is given as:

\begin{eqnarray}
 C_\upsilon = \dfrac{1}{k_BT^2} \Biggl\{ -\Bigl( \dfrac{1}{Z} \sum_i E_i\exp{(-\beta E_i)} \Bigl)^2 + \dfrac{1}{Z}\sum_iE_i^2\exp{(-\beta E_i)} \Biggl\}
\label{Cv}
\end{eqnarray}

where $\beta=1/k_BT$. 

\begin{figure}[t] 
    \centering
    \hspace*{-0.0in}
    \includegraphics[width=0.48\textwidth]{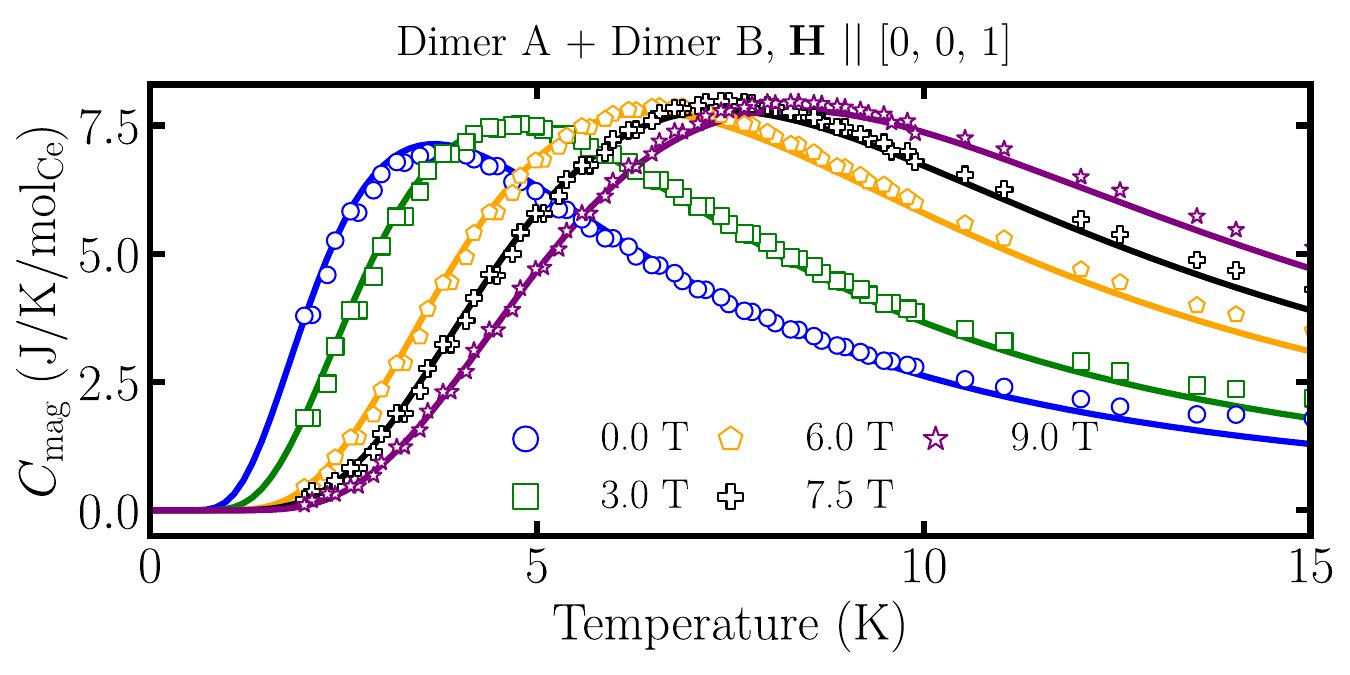}
    \caption{Experimental and calculated specific heat with magnetic field along c-axis. The discrete data points are measured from experiments and the solid lines are calculated results.}
    \label{SI-cp001}
\end{figure}

The in-plane specific heat data has been shown and discussed in main manuscript. Here we show additional specific heat data with external magnetic field applied along c-axis in Fig.~\ref{SI-cp001}. Specific heat along c-axis is most sensitive to the $g_{zz}$. As can be seen, the localized dimer model continues to show excellent agreement with experimental data. The fitting of magnetization, specific heat further refines the exchange matrix components and the anisotropy g-tensors.

\section{Extracting Energy Levels of the Spin Excitations}

\begin{figure}[t] 
    \centering
    \hspace*{-0.0in}
    \includegraphics[width=0.48\textwidth]{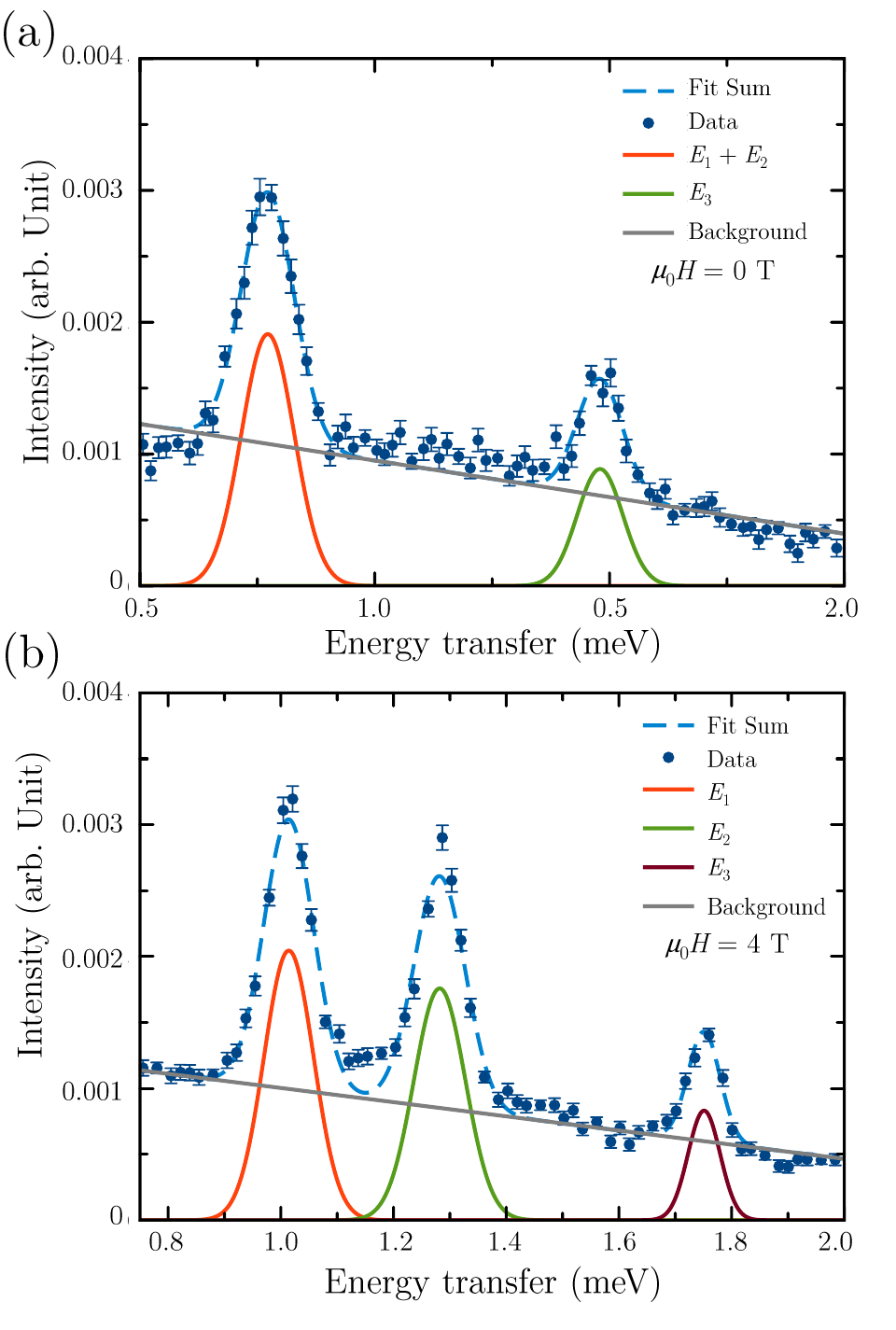}
    \caption{(a) The Gaussian fits and line cuts of intensity vs energy transfer for zero-field data. The data is integrated over $h = [-2,-1]$, $l = [-1.5, 1.5]$ and $\pm0.2$\,$\AA^{-1}$ in the traverse direction. (b) The Gaussian fits and line cuts of intensity vs energy transfer for 4 T data. The data is integrated over $hh = [-1.5,-0.5]\,\AA^{-1}$, $l = [-1.5, 1.5]\,\AA^{-1}$ and $\pm0.2$\,$\AA^{-1}$ in the traverse direction. The integration range in $l$ and the traverse directions are identical to Fig.~2 in the main manuscript.}
    \label{SI-1d}
\end{figure}

The energy levels for the 5 observed quasi-dispersionless spin excitations in zero-field and 4 T are used in determining the intra-dimer exchange matrix constants. The precise energy locations of these spin excitations are extracted by fitting 1-D line cuts on the inelastic neutron scattering (INS) data shown in Fig.~2 (a) and (e) in the main manuscript. The data were fitted with a flat background and Gaussian functions to extract the precise peak positions (in meV) and peak width (HWHM). The fitted results are shown in Fig.~\ref{SI-1d}. Panel (a) shows the line cut for zero-field data by integrating over $h = [-2,-1]\,\AA^{-1}$, $l = [-1.5, 1.5]\,\AA^{-1}$ and $\pm0.2$\,$\AA^{-1}$ in the traverse direction. Similarly the 4 T line cut result is shown in Fig.~\ref{SI-1d} (b). The extracted energy levels and their respective peak widths are summarized in Table.~S\ref{SI-1d-table}.

\begin{table}[]
    \begin{tabular}{|c|c|c|c|c|}
    \hline
                      &  0 T   & HWHM &  4 T   & HWHM \\
    \hline
        $E_1$ (meV) & 0.77 (1) & 0.066(4)& 1.01 (1) & 0.052(3)\\
    \hline
        $E_2$ (meV) & 0.77 (1) & 0.066(4)& 1.28 (1) & 0.054(3)\\
    \hline
        $E_2$ (meV) & 1.48 (1) & 0.057(6)& 1.75 (1) & 0.033(4)\\
    \hline
    \end{tabular}
    \caption{Summary of the fitted energy levels (in meV) and the peak width(HWHM) for zero-field and 4 T INS data obtained from CNCS.}
    \label{SI-1d-table}
\end{table}